\documentclass[twocolumn,preprintnumbers,aps,prl,superscriptaddress,nofootinbib,floatfix,showpacs,showkeys]{revtex4-1}

\usepackage{braket}
\usepackage{float}
\usepackage{graphicx}  
\usepackage{csquotes}
\usepackage{multirow}
\usepackage{natbib}
\usepackage{pifont}
\usepackage{amssymb}
\usepackage{bm}        
\usepackage{amsfonts}  
\usepackage{amsmath}   
\usepackage{amssymb}   
\usepackage[titletoc]{appendix}
\usepackage{color}
\usepackage{hyperref}
\usepackage{cleveref}
\usepackage[english]{babel}
\usepackage{booktabs}
\usepackage{mathtools}
\usepackage{subfigure}
\usepackage{enumerate}

\definecolor{vividviolet}{rgb}{0.62, 0.0, 1.0}
\definecolor{amaranth}{rgb}{0.9, 0.17, 0.31}
\definecolor{palatinateblue}{rgb}{0.15, 0.23, 0.89}
\definecolor{brightpink}{rgb}{1.0, 0.0, 0.5}
\definecolor{cornflowerblue}{rgb}{0.39, 0.58, 0.93}
\definecolor{deepcarminepink}{rgb}{0.94, 0.19, 0.22}
\definecolor{radicalred}{rgb}{1.0, 0.21, 0.37}

\hypersetup{ linktoc=all,
    colorlinks, linkcolor={palatinateblue},
    citecolor={brightpink}, urlcolor={amaranth}
}

\def\be{\begin{equation}}
\def\ee{\end{equation}}

\begin{document}

\title{Geometric multipartite entanglement from gravitational particle production}

\author{Alessio Belfiglio}
\email{alessio.belfiglio@unisi.it}
\affiliation{DSFTA, University of Siena, Via Roma 56, 53100 Siena, Italy.}

\author{Roberto Franzosi}
\email{roberto.franzosi@unisi.it}
\affiliation{DSFTA, University of Siena, Via Roma 56, 53100 Siena, Italy.}
\affiliation{Istituto Nazionale di Fisica Nucleare (INFN), Sezione di Perugia, Perugia, 06123, Italy.}

\author{Orlando Luongo}
\email{orlando.luongo@unicam.it}
\affiliation{University of Camerino, Via Madonna delle Carceri, Camerino, 62032, Italy.}
\affiliation{Department of Mathematics and Physics, SUNY Polytechnic Institute, Utica, NY 13502, USA.}
\affiliation{INAF - Osservatorio Astronomico di Brera, Milano, Italy.}
\affiliation{Al-Farabi Kazakh National University, Almaty, 050040, Kazakhstan.}

\begin{abstract} 
We explore novel generation of genuine multipartite entanglement within gravitational particle production processes during inflationary stages. To this end, we focus on perturbative production mechanisms, considering a non-minimally coupled scalar inflaton field with quartic self-coupling potential and computing probability amplitudes arising from its gravitational interaction with background perturbations. The corresponding entanglement amount is quantified using the recently proposed Entanglement Distance, that provides a \emph{geometric interpretation of particle entanglement, in terms of the Fubini-Study metric}. We observe that, in the limit of negligible squeezing, the total amount of entanglement is dominated by the infrared cutoff scale, in agreement with previous studies analyzing the von Neumann entropy within bipartite scenarios. We then show that \emph{non-negligible multipartite entanglement signatures may emerge across inflation, even during the latest stages of slow-roll}, highlighting their dependence on inflationary momentum scales. Generalizations to regimes with non-negligible squeezing, cubic non-Gaussianities, additional spectator fields and possible observational signatures are also discussed.
\end{abstract}

\pacs{03.67.Bg, 03.67.Mn, 04.62.+v, 98.80.Cq}

\maketitle 
%\tableofcontents
%%%%%%%%%%%%%%%%%%%%%%%%%%%%%%%%%%%%%%%%%%%%%%%%%%%%%%%%%%%%%%%%%%%%%%%

\section{Introduction}

Gravitational particle production (GPP) \cite{PhysRevLett.21.562,PhysRev.183.1057,PhysRevD.3.346,RevModPhys.96.045005} is currently attracting significant attention, in view of tracing back dark matter generation to purely gravitational mechanisms \cite{Kolb:1998ki,PhysRevD.64.043503,Kolb:2007vd, PhysRevD.101.083516, Cembranos:2019qlm, PhysRevD.110.023541}. Remarkably, GPP has also shown impact within reheating \cite{PhysRevD.99.043008,PhysRevD.101.063529} and baryogenesis \cite{Bambi:2006hp,PhysRevD.107.063537} phases, and it may also have implications for primordial black hole formation \cite{Erfani:2015rqv}. 

The overall mechanism  generally arises from the combined effects of cosmological expansion and spacetime inhomogeneities, allowing energy and momentum transfer from the (classical) background gravitational field to quantum fields. Initial studies on GPP were performed by assuming an unperturbed background expansion, typically modeled by a Friedmann-Robertson-Walker  spacetime, and then computing the quantum field modes in the asymptotic past and future, related by appropriate Bogoliubov transformations \cite{Parker:2009uva}.

However, the additional presence of inhomogeneities is able to \emph{enhance} the GPP mechanism from vacuum \cite{PhysRevD.39.389,PhysRevD.45.4428}, leading to possibly larger number densities and introducing mode-mixing in particle creation processes. Further, in some early-time scenarios, the dominant contribution to GPP is expected to directly arise from spacetime perturbations \cite{Bassett:2001jg,Belfiglio:2025chv}.

Quite importantly, \emph{GPP processes may also generate entanglement}, in the final state of the system. In the limit of negligible inhomogeneities, spacetime expansion simply provides particle-antiparticle pairs. The corresponding entanglement entropy is then quantified via the standard von Neumann entropy of the reduced density operator,  tracing out the particle or antiparticle contribution \cite{Ball:2005xa,Fuentes:2010dt,Moradi:2013uya}. However, the additional presence of perturbations modifies this scenario \cite{Belfiglio:2022cnd,Belfiglio:2022yvs}, \emph{allowing for multiparticle production and thus enriching the corresponding entanglement structure}.

Accordingly, a proper characterization of multipartite entanglement in cosmological settings is expected to have profound consequences from a theoretical, observational and experimental point of view, turning out to be fully-unexplored to date. Indeed, entanglement entropy plays a key role in addressing the quantum-to-classical transition of cosmological perturbations \cite{Kiefer:1998qe,Kiefer:2008ku}, where squeezing \cite{PhysRevD.42.3413,Gasperini:1993mq,Agullo:2022ttg} and decoherence \cite{Martineau:2006ki,Nelson:2016kjm,Brahma:2021mng, Salcedo:2024smn} are both deeply connected to GPP mechanisms even beyond the quadratic order \cite{PhysRevD.102.043529,PhysRevD.108.043522}. 

Moreover, the potential detectability of primordial non-Gaussianities may result in additional hints towards the quantum origin of the cosmic microwave background (CMB) radiation \cite{Brahma:2022yxu,PhysRevD.110.043512}, \emph{de facto} permitting to discriminate different inflationary models \cite{PhysRevD.84.083504}.  At the same time, the possibility to simulate GPP processes via laboratory analogues \cite{Barcelo:2005fc} may provide entanglement resources directly usable for quantum information purposes \cite{Steinhauer:2015saa,PhysRevLett.128.091301,PhysRevD.106.105021,PhysRevA.109.013305}.

In this work, we focus on the emergence of multipartite entanglement signatures from GPP processes during the early stages of the Universe's evolution. In particular, we work out entanglement generation associated with the quantum fluctuations of a non-minimally coupled scalar inflaton field. In addition to the standard coupling term between the inflaton and the scalar curvature of spacetime, which can be relevant in framing both inflationary and post-inflationary dynamics \cite{Hertzberg:2010dc,Belfiglio:2024swy,Luongo:2024opv}, we select a quartic self-coupling potential, resulting in one of the most promising single-field inflationary models \cite{Planck:2018jri}. We then evaluate GPP probability amplitudes due to inflaton-induced background perturbations \cite{PhysRevD.107.103512,PhysRevD.109.123520}, focusing on fluctuation modes which exhibit negligible squeezing. 

The corresponding entanglement amount is quantified via the Entanglement Distance (ED) \cite{PhysRevA.101.042129}, a recently proposed entanglement monotone, arising from an adapted application of the \emph{Fubini-Study metric} \cite{Gibbons:1991sa,PhysRevLett.72.3439} and currently investigated in various quantum key distribution protocols for both \emph{qubit} and \emph{qudit} applications \cite{Levay:2003pb,PhysRevA.80.042302,PhysRevA.95.042333,PhysRevX.9.041042}. The geometric nature of the ED in multiqubit scenarios, associated with the distance between infinitesimally close states \cite{Vesperini:2023wks}, provides a natural framework to understand how particle entanglement responds to the underlying spacetime geometry. Hence, quite remarkably, \emph{our outcomes reveal a rich entanglement structure beyond standard bipartite approaches}, also outlining a marked sensitivity of the ED to the inflationary infrared cutoff scale. Thus, while these results are consistent with previous findings involving the von Neumann entropy, the ED is able to capture additional entanglement features arising from   multiparticle GPP processes. Furthermore, we discuss extensions to non-vanishing squeezing regimes, including cubic-order perturbative corrections, thereby opening new avenues toward multipartite entanglement generation at early times. These may result in non-negligible observational signatures, especially in view of entanglement harvesting protocols \cite{Reznik:2002fz,PhysRevD.79.044027,PhysRevD.92.064042}, delineating new pathways for empirical investigation in upcoming experimental programs.

\section{Inflationary setup}

We consider a non-minimally coupled scalar inflaton field, $\phi$, with corresponding Lagrangian density
\begin{equation}\label{lagrinf}
    \mathcal{L}_I=\frac{1}{2} \left[ g^{\mu \nu} \phi_{, \mu} \phi_{,\nu}- \xi R \phi^2 \right]- V(\phi),
\end{equation}
where $\xi$ is the field-curvature coupling constant and we set $V(\phi)= \lambda \phi^4/4$, resulting in one of the most likely single-field inflationary models, according to Planck data \cite{Planck:2018jri}. The universe expansion during inflation can be modeled by a spatially flat Friedmann-Robertson-Walker background, whose line element in cosmic time $t$ reads
$ ds^2=dt^2-a^2(t) d{\bf x}^2$.

At this stage, we move to \emph{conformal time}, $\tau= \int dt/a(t)$, in order to simplify the inflationary dynamics during slow-roll and, therefore, we write the unperturbed metric as
$g_{\mu \nu}= a^2(\tau) \eta_{\mu \nu}$, 
where $\eta_{\mu \nu}$ describes Minkowski spacetime.

The inflaton quantum fluctuations are introduced according to ansatz $\phi({\bf x},\tau)=\phi_0(\tau)+ \delta \phi ({\bf x},\tau)$, where the background homogeneous contribution $\phi_0$ has been isolated from the perturbing quantum fluctuations $\delta \phi$. To linear order, the most general metric tensor describing scalar perturbations can be expressed in the form
\be \label{metrper}
g_{\mu \nu}= a^2(\tau) \begin{pmatrix}
    1+2\Phi & \partial_i B \\
    \partial_i B & -\left( (1-2\Psi)\delta_{ij}+ D_{ij} E  \right) \\
\end{pmatrix},
\ee
where $\Phi$, $\Psi$, $B$ and $E$ are scalar quantities and $D_{ij}\equiv \partial_i \partial_j - \frac{1}{3} \delta_{ij} \nabla^2$. In order to further account for slow-roll, we select a quasi-de Sitter scale factor,
\be \label{quasids}
a(\tau)= -\frac{1}{H_I \left( \tau-2\tau_f\right)^{(1+\epsilon)}},\ \ \ \ \epsilon \ll 1,
\ee
where $\tau_f$ indicates the end of inflation, $H_I$ is the \emph{Hubble parameter during slow-roll}, while $\epsilon$ denotes a small and constant \emph{slow-roll parameter}. The latter can be quantified via \cite{Riotto:2002yw,Planck:2018jri},
\be \label{scapow}
\epsilon= \frac{1}{\pi \mathcal P_s} \left( \frac{H_I}{M_{\rm pl}}  \right)^2
\ee
where $M_{\rm pl}$ is the Planck mass and $\mathcal P_s$ is the dimensionless scalar power spectrum, observationally constrained at $\mathcal P_s=2.1 \times 10^{-9}$ for the pivot scale $k_{\rm piv}=0.05$ Mpc$^{-1}$, see e.g. Ref. \cite{Planck:2018jri}. 

Hereafter, we adopt the \emph{longitudinal}, or conformal Newtonian gauge \cite{Mukhanov:1990me}, corresponding to $E=B=0$. Moreover, when the fluctuation matter source has no anisotropic stress, as for the scalar inflaton, we can further set $\Phi=\Psi$.

Moving to Fourier space, we write perturbation modes as
\be \label{pertmod}
\Psi({\bf x}, \tau)= \frac{1}{\left( 2\pi \right)^{3/2}}\int d^3k\Psi_k(\tau) e^{i {\bf k}\cdot {\bf x}},
\ee
and, similarly, we can expand quantum fluctuations as
\be \label{qfesp}
\hat{\delta \phi} ({\bf x}, \tau)= \frac{1}{(2 \pi)^{3/2}} \int d^3k \left( \hat{a}_k \delta\phi_k e^{i {\bf k}\cdot {\bf x}} + \hat{a}_k^\dagger \delta\phi_k^* e^{-i {\bf k}\cdot {\bf x}} \right), 
\ee
satisfying the usual canonical commutation relations, $
[\hat{a}_k, \hat{a}_{k^\prime}^\dagger]= \delta^{(3)} ({\bf k}-{\bf k^\prime})$. In the limit $\lvert \xi \rvert \ll 1$, the first-order perturbed Einstein equations\footnote{For simplicity, we denote $\phi_0$ by $\phi$ from now on.} give $\Psi_k^\prime+\mathcal{H} \Psi_k= \epsilon \mathcal{H}^2 \delta \phi_k/\phi^\prime$. Conformally rescaling inflaton fluctuations by $\delta \chi_k= \delta \phi_k a$ , we then obtain 
\be \label{modes}
\delta \chi_k^{\prime \prime}+ \left[ k^2-\frac{1}{\eta^2} \left( (1-6\xi)(2+3\epsilon) + 6\epsilon - \frac{ V_{,\phi \phi}}{H_I^2} \right)    \right] \delta \chi_k=0,
\ee
where $k \equiv \lvert {\bf k} \rvert$, $\eta \equiv \tau-2\tau_f$ and we also computed the scalar curvature in conformal time, $R=6a^{\prime \prime}/a^3$, noting that $a^{\prime \prime}/a \simeq (2+3\epsilon)/\eta^2$. Selecting the Bunch-Davies initial conditions \cite{Bunch:1978yq,Danielsson:2003wb,Greene:2005wk}, Eq. \eqref{modes} gives
\be \label{influct}
\delta \chi_k (\eta)= \frac{\sqrt{-\pi\eta}}{2} e^{i\left( \nu+ \frac{1}{2}\right) \frac{\pi}{2}}  H^{\left(1\right)}_{\nu}\left(-k\eta\right),
\ee
where $H_\nu^{(1)}$ denotes the Hankel function of first kind, with
\be \label{hankind}
\nu= \sqrt{1/4+(1-6\xi)(2+3\epsilon) + 6\epsilon-V_{,\phi \phi}/H_I^2}
\ee
the corresponding index.

\section{Perturbative GPP and multipartite entanglement}

Starting from Eq. \eqref{metrper} and resorting to the longitudinal gauge, we can write $g_{\mu \nu}= a^2(\tau) \left( \eta_{\mu \nu}+ h_{\mu \nu} \right)$, with $h_{\mu \nu}=\text{diag}\left( 2\Psi, 2\Psi, 2\Psi, 2\Psi  \right)$.

The first-order interaction Lagrangian density describing the coupling between inflaton fluctuations and metric perturbations then reads \cite{PhysRevD.39.389}
\be \label{intlag}
    \mathcal{L}_{I}=-\frac{1}{2}\sqrt{-g_{(0)}}H^{\mu\nu}T^{\left(0\right)}_{\mu\nu},
\ee
where $H_{\mu \nu}= a^2(\tau) h_{\mu \nu}$, $g_{(0)}$ is the determinant of the background unperturbed metric tensor and $T_{\mu \nu}^{(0)}$ the energy-momentum tensor for fluctuations computed at zero order \cite{Riotto:2002yw}, i.e., neglecting additional backreaction effects of metric perturbations on the inflaton field.

For a Lagrangian density as in Eq. \eqref{intlag}, we can write the $\hat{S}$ matrix at first order using Dyson's expansion as $\hat{S} \simeq 1 + i \hat{T} \int d^4x \mathcal{L}_I$. In the limit $\lvert \xi \rvert \ll 1$, perturbative particle production during slow-roll is governed by $V(\phi)$, giving
\begin{align} \label{probamp_gen}
\mathcal{C}_{{\bf k}_1,{\bf k}_2,{\bf k}_3,{\bf k}_4} & \equiv \langle {\bf k}_1, {\bf k}_2, {\bf k}_3, {\bf k}_4 \lvert \hat{S} \rvert 0 \rangle \notag \\[3pt]
& =  \frac{4! \ i \lambda}{2(2\pi)^6} \int d^4x\   \Psi a^4  \delta \phi^*_{k_1} \delta \phi^*_{k_2} \delta \phi^*_{k_3} \delta \phi^*_{k_4} \notag \\
&\ \ \ \ \ \ \ \ \ \ \ \ \ \ \ \ \times e^{- i \left({\bf k}_1 + {\bf k}_2 + {\bf k}_3 + {\bf k}_4 \right) \cdot {\bf x} }, 
\end{align}
where we have assumed negligible squeezing\footnote{Generalization to nonzero squeezing would result in additional contributions to the total probability amplitude. See e.g. Ref. \cite{PhysRevD.102.043529} for an application of perturbation theory to squeezed cubic non-Gaussianities.} for all the involved modes.

From Eq. \eqref{pertmod}, we readily obtain
\begin{align} \label{probamp}
   \mathcal{C}_{{\bf k}_1,{\bf k}_2,{\bf k}_3,{\bf k}_4} \propto \int d\tau  a^4 \Psi_{\lvert {\bf k}_1+{\bf k}_2+{\bf k}_3+{\bf k}_4 \rvert} \delta \phi^*_{k_1} \delta \phi^*_{k_2} \delta \phi^*_{k_3} \delta \phi^*_{k_4},
\end{align}
namely the background gravitational field is able to transfer momentum to the field modes $\delta \phi_k$.

\subsection{Entanglement Distance from generalized Fubini-Study metric}

Let us consider a single production event, corresponding to the final state
\be \label{finstate}
\ket{\Phi}= N \left( \ket{0000} + \frac{\mathcal{C}_{{\bf k}_1,{\bf k}_2,{\bf k}_3,{\bf k}_4}}{4!} \ket{1111}  \right),
\ee
where $N$ denotes a normalization constant. Eq. \eqref{finstate} displays a Greenberger-Horne-Zeilinger (GHZ)-like state, widely employed in quantum information protocols, exhibiting genuine multipartite entanglement \cite{PhysRevA.63.012308,PhysRevLett.98.070502}. 

\begin{figure}
    \centering
    \includegraphics[scale=0.65]{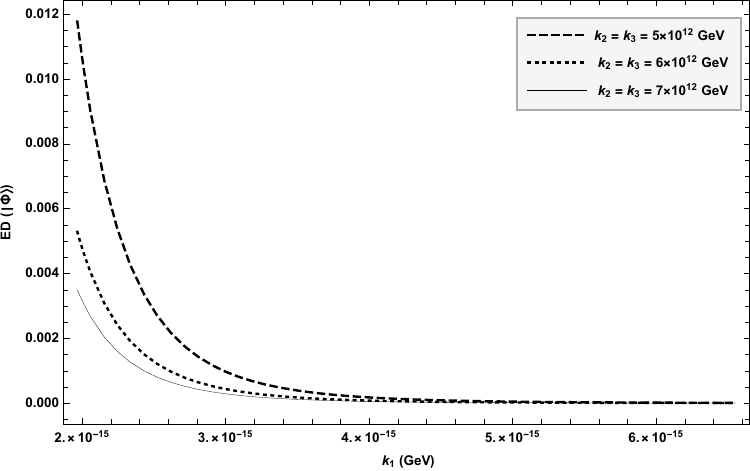}
        \caption{Entanglement Distance for a perturbative gravitational quartet production process. Here, one excitation (labeled by ${\bf k}_1$) is produced on super-Hubble scales, while the remaining three have sub-Hubble momenta. We set $k_2 = k_3 = k_4/2 \gtrsim H_I$ and ${\bf k}_2 + {\bf k}_3 + {\bf k}_4 = 0$, in such a way to preserve the classicality condition for perturbation modes $\Psi_k$. The probability amplitude $\mathcal{C}_{{\bf k}_1,{\bf k}_2,{\bf k}_3,{\bf k}_4}$ is computed focusing on the last inflationary e-folding $[\tau_1, \tau_f]$, with $a(\tau_1) H_I/1000 < k_1 < a(\tau_1) H_I/100$. The other parameters are: $\lambda=10^{-16}$ and $H_I \simeq 4.5 \times 10^{12}$ GeV, corresponding to $\phi(\tau_{\rm piv})= 5\ M_{\rm pl}$.}
    \label{fig_entad}
\end{figure}

The corresponding Entanglement Distance can be computed from the general formula 
\be \label{ed_def} 
ED\left(\ket{\Phi}\right)= D- \sum_{\mu=0}^{D-1} || \bra{\Phi} \boldsymbol{\sigma}^\mu \ket{\Phi}    ||^2,
\ee
where $D$ is the total number of qubits and $\boldsymbol{\sigma}^\mu$ is the vector of Pauli matrices acting on the $\mu$-th qubit, namely $\boldsymbol{\sigma}^\mu= \left( \sigma^\mu_x, \sigma^\mu_y, \sigma^\mu_z \right)$. In the case of multiqubit states, the ED can be interpreted as an obstacle to the minimum distance between infinitesimally close states, according to the Fubini-Study metric of the projective Hilbert space associated with the quantum system \cite{PhysRevA.101.042129,Vesperini:2023wks}.

From Eq. \eqref{finstate}, we observe that only $\sigma^\mu_z$ contributes to the above sum, for each $\mu$. In particular, defining 
\be \label{redop_1}
\rho_\mu= \text{Tr}_{\nu \neq \mu}\left( \ket{\Phi} \bra{\Psi} \right),
\ee
we find 
\begin{align} \label{expval_z}
\text{Tr}\left( \rho_\mu \sigma^\mu_z \right) &= \text{Tr} \left( N^2\begin{pmatrix}
    1 & 0 \\ 0 &  \ \lvert \mathcal{C}_{{\bf k}_1,{\bf k}_2,{\bf k}_3,{\bf k}_4} \rvert^2
\end{pmatrix}   \begin{pmatrix}
    1 & 0\\ 0 & -1
\end{pmatrix}   \right) \notag \\[3pt]
&= N^2 \left( 1- \lvert \mathcal{C}_{{\bf k}_1,{\bf k}_2,{\bf k}_3,{\bf k}_4} \rvert^2    \right),
\end{align}
where, for simplicity, we have rescaled the probability amplitude by $\mathcal{C}_{{\bf k}_1,{\bf k}_2,{\bf k}_3,{\bf k}_4}/4! \rightarrow \mathcal{C}_{{\bf k}_1,{\bf k}_2,{\bf k}_3,{\bf k}_4}$. Similarly, one finds $\text{Tr}\left( \rho_\mu \sigma^\mu_x \right) = \text{Tr}\left( \rho_\mu \sigma^\mu_y \right)=0$.  The ED then reads
\be \label{ed_comp}
ED\left( \ket{\Phi} \right) =  4- 4 \left[ N^2 \left(  1- \lvert \mathcal{C}_{{\bf k}_1,{\bf k}_2,{\bf k}_3,{\bf k}_4} \rvert^2 \right)   \right]^2,
\ee
corresponding to the Fubini-Study entanglement metric \cite{PhysRevA.101.042129}
\be \label{entamet}
\tilde{g} = \left( 1- 1 \left[ N^2 \left(  1- \lvert \mathcal{C}_{{\bf k}_1,{\bf k}_2,{\bf k}_3,{\bf k}_4} \rvert^2 \right)   \right]^2    \right) J_4,
\ee
where $J_4$ is the $4 \times 4$ matrix containing all ones. The off-diagonal elements of $\tilde{g}$ provide the quantum correlations between qubits. Furthermore, states differing from one another via local unitary transformations have the same form of $\tilde{g}$. Eq. \eqref{entamet} then confirms the \emph{global} nature of the entanglement associated with the state in Eq. \eqref{finstate}, which cannot be captured by standard bipartite measures \cite{PhysRevD.102.043529,Belfiglio:2022yvs}.

\subsection{Observational consequences of cosmic multipartite entanglement}

In Fig. \ref{fig_entad}, we show the mode dependence of the ED in the case of a super-horizon creation process with $k_1 \lesssim aH_I$ and $k_2,k_3,k_4 \gg aH_I$, further assuming ${\bf k}_2 + {\bf k}_3 + {\bf k}_4 = 0$, which allows to preserve the classicality \cite{Bassett:2001jg} of the corresponding perturbation mode $\Phi_k$. We select $N \equiv \log\left[ a(\tau_f)/a(\tau_i) \right] = 70$ inflationary e-foldings, where the inflationary onset, $\tau_i$, is computed from the condition $N-N_{\rm piv} \equiv \log\left[ a(\tau_{\rm piv})/a(\tau_i) \right] =5$, with $\tau_{\rm piv}$ obtained from the standard horizon crossing condition $k_{\rm piv}\equiv a(\tau_{\rm piv})H_I$.

We observe that the ED is larger at small momenta, reflecting the bosonic nature of the involved inflaton field. The same conclusion has been found in the case of non-perturbative gravitational production \cite{Ball:2005xa} and for perturbative pair production processes \cite{PhysRevD.109.123520}. This confirms that the total entanglement is expected to increase in the vicinity of the infrared cutoff, even when dealing with multiparticle processes. However, the presence of large squeezing may change this scenario, as already found in Ref. \cite{PhysRevD.102.043529} for the case of bipartite entanglement, by consistently tracing out sub-Hubble inflationary modes $k \gg aH_I$.
%%%%%%%%%%%%%%%%%%%%%%%%%%%%%%%%%%%%%%%%%%%%%%%%%%%%%%%%%%%%%%%%%%%%%%%%%%%
\begin{figure}
    \centering
    \includegraphics[scale=0.83]{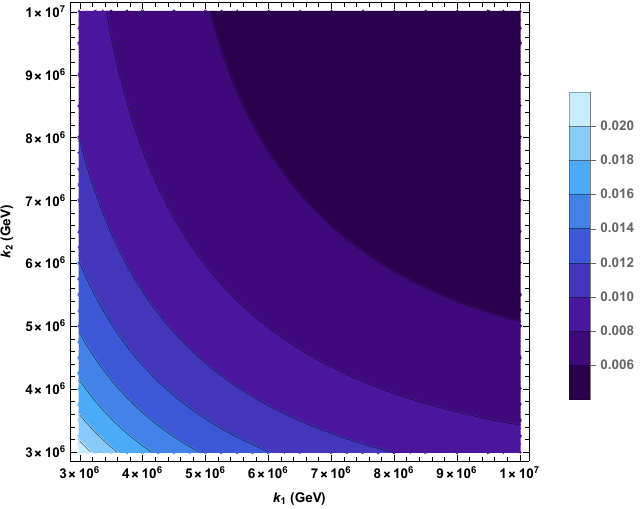}
    \caption{Entanglement Distance in the case of sub-horizon quartet production, assuming $\lvert {\bf k}_1 + {\bf k}_2 + {\bf k}_3 + {\bf k}_4 \rvert = a(\tau_1)H_I/1000$. For simplicity, we select $k_4 \simeq k_1+k_2+k_3= 3\times 10^7$ GeV and we focus again on the latest slow-roll stage of inflation, namely $\tau \in \left[ \tau_1, \tau_f \right]$. The other parameters are the same as in Fig. \ref{fig_entad}.}
    \label{fig_entad_contour}
\end{figure}

%%%%%%%%%%%%%%%%%%%%%%%%%%%%%%%%%%%%%%%%%%%%%%%%%%%%%%%%%%%%%%%%%%%%%%%%%%%

The scale dependence of the ED is preserved in the case of sub-horizon production processes, as we show in Fig. \ref{fig_entad_contour}. Within this framework, we retain the classicality of $\psi_k$ by appropriately selecting $k_1,k_2,k_3,k_4 \gg aH_I$, thus capturing entanglement features that cannot be described in standard bipartite scenarios, where either sub- or super-Hubble modes are necessarily traced out. The ED is again modulated by the perturbation mode $\Psi_k$, confirming, at the same time, that is typically harder to entangle scalar particles with larger momenta.  

Sub-horizon entanglement generation from GPP may provide relevant information concerning the microphysics of reheating and the early stages of the radiation phase, both for primordial fluctuations and inflationary spectator fields, which are typically assumed to weakly interact with the quantum fields in the standard model of particle physics and \emph{may thus represent plausible dark matter candidates}, see e.g. Ref. \cite{RevModPhys.96.045005} for a recent review.

\section{Final outlooks}

In this work, we investigated multipartite entanglement generation arising from GPP processes during inflation. For conceptual clarity, we adopted a simplified setup, already studied in recent literature, according to which the scalar inflaton fluctuations interacts with metric perturbations generated by the inflaton field itself. The entanglement amount is quantified via the ED, which, in the case of multiqubit states, can be interpreted as the minimum distance between infinitesimally closed states, arising from the Fubini-Study metric. 

\emph{Our outcomes show the presence of genuine multipartite entanglement features within realistic inflationary scenarios}, which cannot be captured by standard bipartite approaches.

In particular, assuming negligible squeezing, i.e., neglecting modes that leave the Hubble horizon several e-foldings before the end of inflation, the choice of a self-coupling quartic inflaton potential may lead to quadripartite GHZ-like states. The corresponding entanglement amount is sensitive to the infrared cutoff and it is typically larger at smaller momenta, in agreement with the von Neumann entropy of previously studied bipartite particle states.
The presence of large squeezing, which characterized, for example, the pivot scales currently probed by the Planck satellite, would enrich the above-presented picture, by providing additional contributions to the total probability amplitude for GPP. In addition, the squeezing of inflationary fluctuations may affect the ED scaling with respect to the momentum of the involved particles.

Last but not least, multipartite entanglement measures are fundamental in addressing the dynamics of cubic, and/or higher order, non-Gaussianities on the CMB, whose entanglement features cannot be fully described by the von Neumann entropy. A gauge-invariant approach to the entanglement of such contributions will shed further light on the quantum-to-classical transition of cosmological perturbations, which has received significant attention in recent years. 

Particularly, a multipartite approach will allow to distinguish entanglement features both on sub and super-Hubble scales, with important consequences not only for CMB observations, but also for the latest stages of inflation and reheating. 

In this direction, \emph{the ED would adapt as well to the study of perturbative GPP associated with spectator fields}, thus allowing to better understand the quantum properties of the corresponding particles, which may, for example, represent plausible dark matter candidates.

\section*{Acknowledgements}
OL acknowledges financial support from the Fondazione  ICSC, Spoke 3 Astrophysics and Cosmos Observations. National Recovery and Resilience Plan (Piano Nazionale di Ripresa e Resilienza, PNRR) Project ID CN$\_$00000013 "Italian Research Center on  High-Performance Computing, Big Data and Quantum Computing"  funded by MUR Missione 4 Componente 2 Investimento 1.4: Potenziamento strutture di ricerca e creazione di "campioni nazionali di R$\&$S (M4C2-19 )" - Next Generation EU (NGEU)
GRAB-IT Project, PNRR Cascade Funding
Call, Spoke 3, INAF Italian National Institute for Astrophysics, Project code CN00000013, Project Code (CUP): C53C22000350006, cost center STI442016. 

RF acknowledges the support of the Research Support Plan 2022 - Call for applications for funding allocation to research projects curiosity-driven (F CUR) - Project \enquote{Entanglement Protection of Qubits' Dynamics in a Cavity} - EPQDC. AB and RF would like to acknowledge INFN Pisa for the financial support to this activity.

\end{document}